\documentclass{poster18}

\usepackage[nolist,nohyperlinks]{acronym} % Abkürzungen
\usepackage{textcomp} % textmu

\usepackage{color, colortbl}
\definecolor{Gray}{gray}{0.9}
\definecolor{White}{gray}{1.0}
\usepackage{tabularx}

\begin{document}

		\begin{acronym}[ticktocktick]
		\acro{BA}{Barabasi--Albert}
		\acro{ER}{Erdos--Renyi}
		\acro{CM}[CM]{Connectivity Matrix}	
		\acro{DM}{delay matrix}	
		\acro{FLOP}{float operation}
		\acro{FS}{fast spiking}
		\acro{HDMEA}{High Density Microelectrode Array} 
		\acro{HH}{Hodgkin--Huxley}	
		\acro{IF}{integrate--and--fire}	
		\acro{IC}{implementation of Catanzaro}	
		\acro{SII}{standard implementation of Izhikevich}	
		\acro{MEA}{microelectrode array} 
		\acro{MFR}{Mean Firing Rate}	
		\acro{RS}{regular spiking}
		\acro{STDP}{spike-timing dependent plasticity}
		\acro{SWM}{synaptic weight matrix}
		\acro{TE}{Transfer Entropy}
		\acro{UAS}{University of Applied Sciences}  
	\end{acronym}
	
%----------------------------------------------------------

%----------------------------------------------------------
%               THIS IS THE PLACE OF THE TITLE
%
\title{Simulation of Large Scale Neural Networks\\ for Evaluation Applications}
%----------------------------------------------------------
%               THIS IS THE PLACE FOR THE AUTHORS NAMES AND THE TITLE FOR HEADINGS
%
\headtitle{S. DE BLASI, SIMULATION OF LARGE SCALE NEURAL NETWORKS FOR EVALUATION APPLICATIONS}
%----------------------------------------------------------
%               THIS IS THE PLACE FOR THE AUTHORS NAMES - ALL AUTHORS MUST HAVE A STUDENT STATUS!!!

%
\author{Stefano DE BLASI\affiliationmark{1}}
%----------------------------------------------------------
%              THIS IS THE PLACE FOR AFFILIATIONS
%
\affiliation{%
\affiliationmark{1} Biomems lab, Faculty of Engineering, \ac{UAS} Aschaffenburg, Germany \\
}
  \email{stefano.de.blasi@live.de}
%--------------------------------------------------------------

\maketitle

%----------------------------------------------------------
%               THIS IS THE PLACE FOR ABSTRACT

\begin{abstract}
Understanding the complexity of biological neural networks like the human brain is one of the scientific challenges of our century. The organization of the brain can be described at different levels, ranging from small neural networks to entire brain regions. Existing methods for the description of functionally or effective connectivity are based on the analysis of relations between the activities of different neural units by detecting correlations or information flow.~This is a crucial step in understanding neural disorders like Alzheimer's disease and their causative factors. 
To evaluate these estimation methods, it is necessary to refer to a neural network with known connectivity, which is typically unknown for natural biological neural networks. 
Therefore, network simulations, also in silico, are available.
In this work, the in silico simulation of large scale neural networks is established and the influence of different topologies on the generated patterns of neuronal signals is investigated. 
The goal is to develop standard evaluation methods for neurocomputational algorithms with a realistic large scale model to enable benchmarking and comparability of different studies.
\end{abstract}

%----------------------------------------------------------
%               THIS IS THE PLACE FOR KEYWORDS

%----------------------------------------------------------
%               HERE WRITE YOUR PAPER

\section{Introduction}
The human brain consists of billions of nerve cells, so-called neurons, and is one of the most exciting mysteries of our time.
Many neuroscientists study nervous systems in order to explain the brain, which is essential for understanding human consciousness, motor control or memory and learning. Since brain diseases are widespread and have devastating effects on our lives~\cite{Olesen.2003}, the need for an improvement of our knowledge about these diseases and their causes is critical in order to develop effective treatment. Furthermore, fundamental understanding of the brain could even help to cure neural diseases like epilepsy, Parkinson's disease or Alzheimer's disease~\cite{Bhatti.2017}. 
To understand neural networks like the brain, it is fundamentally necessary to know the information flow based on certain sequences of action potentials, such as bursts, which are very dense sequences of emitted action potentials. How and where is an information, like interpretation of visual or acoustical sensor input, designed and processed by the brain? Are there similarities with technical networks such as the internet?

\begin{figure}[!bp]
	\begin{center}
		\includegraphics[width=0.42\textwidth]{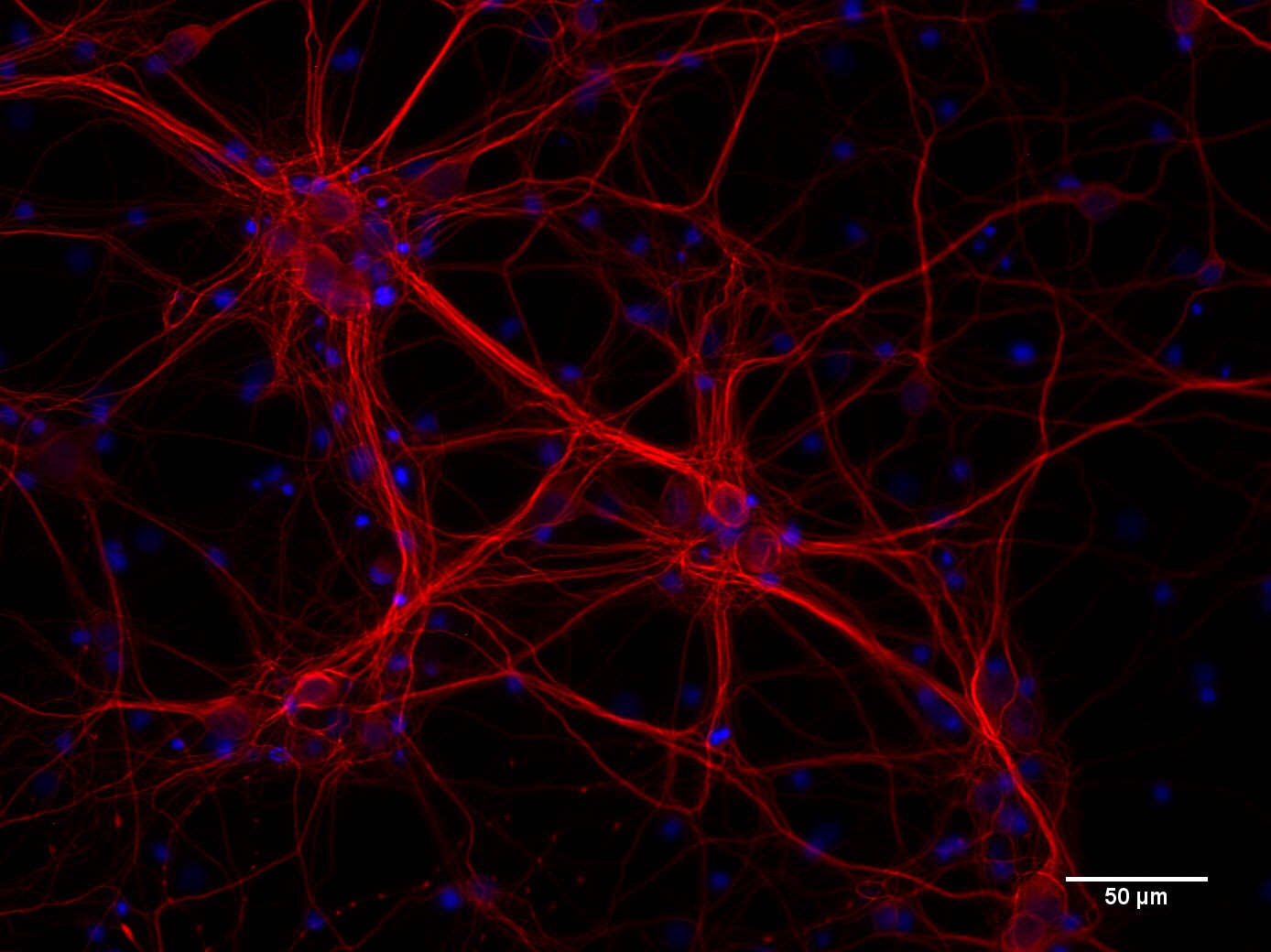}
	\end{center}
	\caption[Biological neural network \textit{in vitro}]{\textbf{Biological neural network  \textit{in vitro}: \\}A fluorescence image of a neuronal \textit{in vitro} culture was taken, with blue coloured cell bodies and red coloured dendrites.  \\}\label{fig:nn}
\end{figure}

Since incoming synaptic action potentials influence the behaviour of the receiving neuron, it is possible to make statements about connections between neurons.
In order to assess the quality of these statements, an evaluation of estimation algorithms requires knowledge about the neural network.
As biological neural networks - even \textit{in vitro} - are far too complex, see Fig.~\ref{fig:nn}, to reveal connections between single neurons, \textit{in silico} neural networks with known topology are typically modelled.
These simulated networks should be biophysically representative for a meaningful evaluation of neurocomputational algorithms, like connectivity estimation, synchrony measurement and burst detection. 

As the topology of an \textit{in silico} network can affect the results and accuracy of algorithms~\cite{Kadirvelu.2017}, it is essential to evaluate these methods with realistic topologies.
In the following, known \textit{in silico} models are compared with respect to their applicability in large-scale simulations for biological neural networks. Implementations for different topologies are analysed in terms of biological plausibility. Based on these findings, recommendations are made for future evaluations of neurocomputational algorithms in order to compare newly developed methods with previous ones in a meaningful way.

\section{Implementation and Modelling}
\subsection*{Neuronal model}
\begin{figure}[bp]
	\begin{center}
		\includegraphics[width=0.34\textwidth]{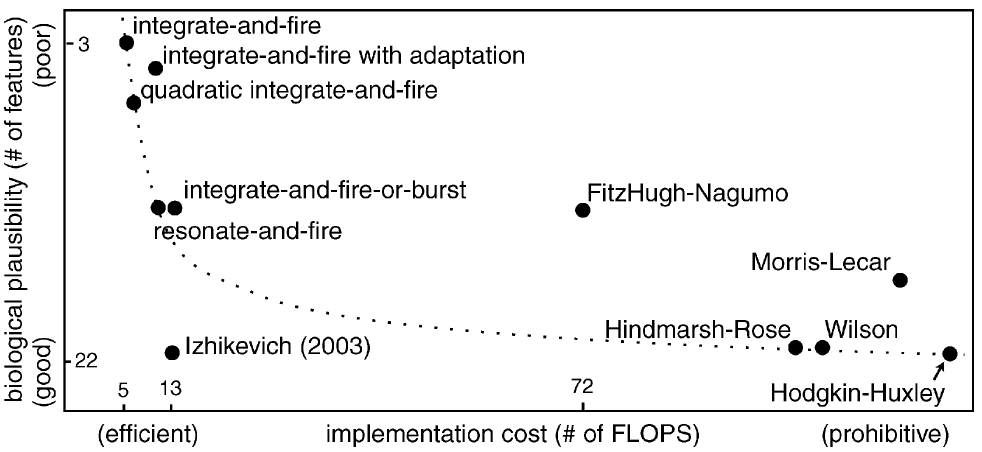}
	\end{center}
	\caption[Comparison of neuronal models]{\textbf{Comparison of neuronal models:\\} Eleven neuronal models are evaluated taking into account the implementation cost in form of \acp{FLOP} and biological plausibility measured by the number of possible features (e.g. tonic spiking, tonic bursting or integrator). Figure and information by Izhikevich~\cite{Izhikevich.2004}.\\}\label{fig:Aihzplot}
\end{figure}

Many neuron models with different strengths and weaknesses were developed, by modelling the synaptic characteristics of inhibitory and excitatory effects. Several models were considered and compared in terms of computing complexity and biological plausibility by investigating their possible features~\cite{Izhikevich.2004}, see Fig.~\ref{fig:Aihzplot}. 
Simulating the well known \ac{HH} models in a realistic number of neurons is computationally intensive. The \ac{IF} model is most efficient for thousands of simulated neurons, but some realistic firing patterns cannot be generated.

The Izhikevich model presents a compromise between the computational efficiency of \ac{IF} model and the biological plausibility of \ac{HH} model, as it is able to generate many firing patterns~\cite{Izhikevich.2004} with only two-differential equations. 
Thus, the model is specially designed for so called large scale simulations, i.e. for up to tens of thousands of neurons, and is able to reproduce spiking and bursting of cortical neurons in real time~\cite{Izhikevich.2003}.

The operating principle of the Izhikevich model bases on the variable $v$, which represents the electrical membrane potential between the interior and the exterior of a neuron, while $u$ describes the membrane recovery variable, which takes the inactivation of sodium $Na^{+}$ and activation of potassium $K^{+}$ channels into account.  Both variables obey the dynamics of membrane potential (\ref{equ:Izh_v}) and recovery (\ref{equ:Izh_u}). Effects driven by incoming synaptic impulses of connected neurons are realized with variable $I$,
\begin{equation}\label{equ:Izh_v}
\dot{v} = 0.04 v^2 + 5 v + 140 - u + I
\end{equation}
\begin{equation}\label{equ:Izh_u}
\dot{u} = a \cdot (b \cdot v - u) .
\end{equation}
Where $a$ is a time scale parameter for $u$, and $b$ describes the sensitivity of $u$ to the subthreshold fluctuations of $v$.
By reaching the threshold of $v \ge 30  \mathrm{~mV} $, the neuron emits an action potential (rapid rise and fall of $v$)  and the auxiliary after-spike resetting is activated. In that case variables $v$ and $u$ are changed obeying the rule (\ref{equ:Izh_aux}). $c$ is the reset potential and parameter $d$ describes the reset value of the recovery variable $u$ after each spike,
\begin{equation}\label{equ:Izh_aux}
\mathrm{~if~}  v \ge 30  \mathrm{~mV, then} \left\{
\begin{array}{ll}
v = c &  \\
u = u + d &  .\\
\end{array}
\right. 
\end{equation}

\subsection*{Neural networks}
Neuronal networks consist of neurons (nodes) and axons, synapses and dendrites (connections), see Fig.~\ref{fig:nn}. The simplest topology is a \textit{regular network} with a constant number of connections, also called degrees, which is exactly the same for each neuron in the whole network. 
A \textit{random network} is constructed by using a constant connection probability following a Poisson distribution,
\begin{equation}\label{equ:rand_d}
P_{(deg = k )} = e ^ {-Np} \cdot \frac{(N \cdot p)^k}{k!} .
\end{equation}

\begin{figure}[!htb]
	\centering
	\includegraphics[width=0.38\textwidth]{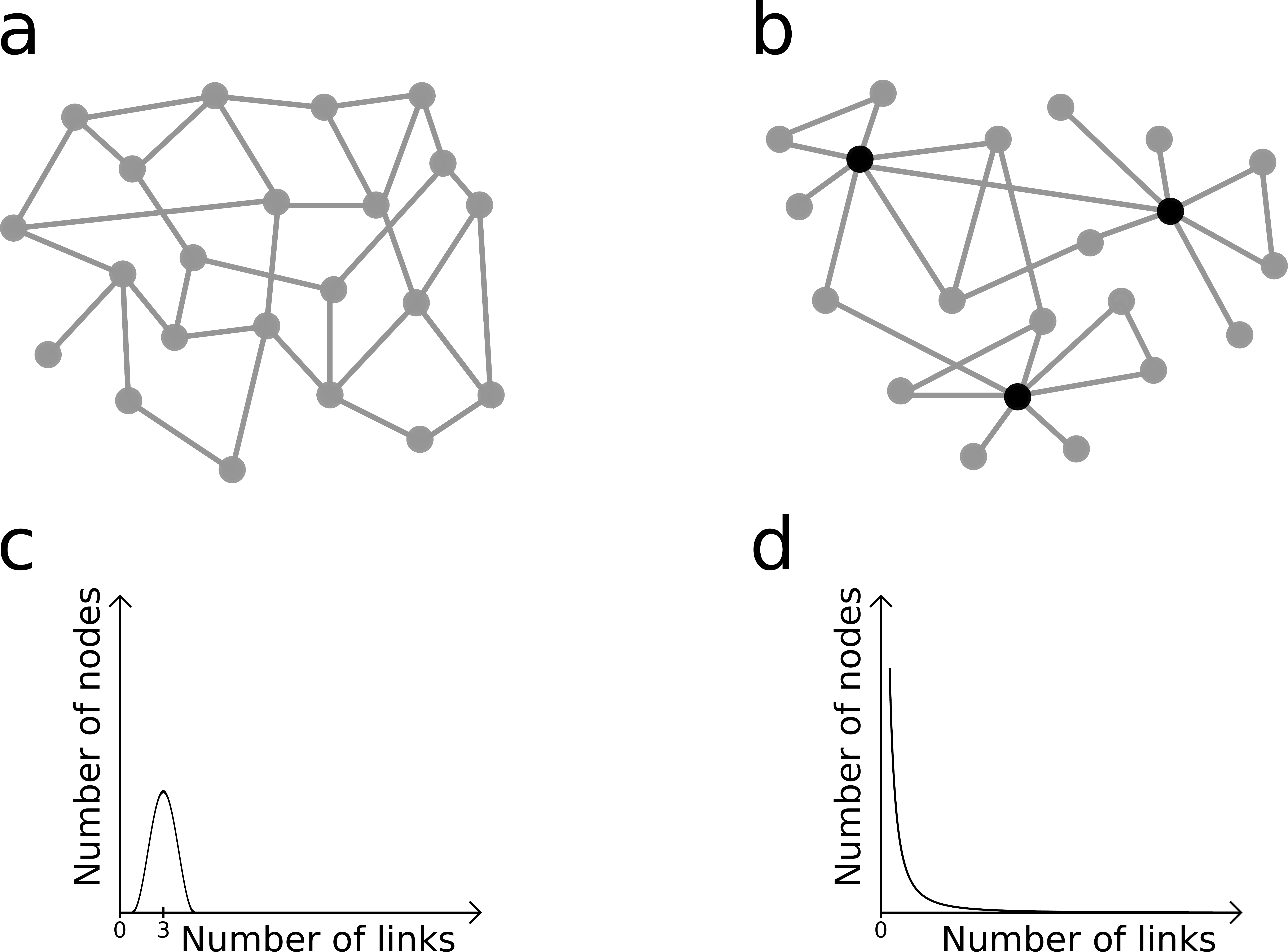}
	\caption[D]{\textbf{Difference between \textit{random} and \textit{scale-free networks}:} 
		A \textit{random network} (a) with mean degree of 3, which is the maximum of its distribution of node linkages (c), and a \textit{scale-free network} (b) with black marked hub neurons are illustrated. For \textit{scale-free networks}, the distribution of node degrees (d) is formed like a power law function.\\}\label{fig:scalefree_vs_rand}
\end{figure}

Nodes with significantly higher or lower numbers of connections are very rare, but unlike the \textit{regular network}, they exist~\cite{Barabasi.2003}. A clearly identifiable mean degree can be recognized for $N \cdot p$, where $N$ is the number of neurons and $p$ is the connection probability. All nodes are connected to the same number of other nodes on average, but there is a standard deviation.
For the illustrated \textit{random network} in Fig.~\ref{fig:scalefree_vs_rand}.a, mean degree would be around three, which is the maximum of degree distribution in Fig.~\ref{fig:scalefree_vs_rand}.c. 
In \textit{scale-free networks} some neurons, so called hubs, have an immense number of connections to other neurons~\cite{Barabasi.2003}. 
As some nodes are barely connected while hubs are able to have 100 times more connections, the \textit{scale-free networks} could also be called ultrasmall-world networks: Some nodes are almost isolated of other groups~\cite{Cohen.2003}.
Hub neurons were already detected in regions of the brain and characterised~\cite{Sporns.2007}.
In Fig.~\ref{fig:scalefree_vs_rand}.b an exemplary \textit{scale-free network} is illustrated with three black marked hub neurons, resulting in a fundamentally different distribution of node degrees, see Fig.~\ref{fig:scalefree_vs_rand}.d. 
This kind of log normal distribution can be described by a power low function with free parameter $\gamma$,
%Thus, there are many nodes which are sparsely connected while some hubs are able to have lots of links.
\begin{equation}\label{equ:sf_d}
P_{(deg = k )} \propto  k^{-\gamma} .
\end{equation}

In this work, a network size of 1000 neurons and the ratio of excitatory to inhibitory neurons of 4 to 1 are chosen, which is standard for Izhikevich simulations. In this model, parameters for \ac{RS} ($a=0.02; b=0.2; c=-65; d=8$) are selected for excitatory neurons and \ac{FS} ($a=0.1; b=0.2; c=-65; d=2$) for inhibitory respectively. While excitatory synapses contribute to the membrane potential of the receiving neuron, inhibitory ones counteract.
For synaptic properties, strengths of connections, so called synaptic weights, are stored in a square symmetric \ac{SWM} with size $N$, which is the number of neurons. Each row represents a target of a synapse. The column index of the \ac{SWM} then indicates the source of this synapse. Value ranges depends on the used model while the sign indicates types of connections where value 0 means unconnected and non-zero values represent the synaptic weights of connections. 
By adaptation of the \ac{SWM}, it is possible to design specific topologies.

\newcolumntype{g}{>{\columncolor{Gray}}c}
\setlength{\arrayrulewidth}{1.0pt}
\begin{table}[!b]
	
	\onecolumn
	\flushleft
	\parbox{1in}{\caption[Summery of network type characteristics]{\textbf{Summery of network type characteristics:\\} Different network characteristics like the existence of hubs, the distribution of input- and output-degree are ensured by various constructional principles, which are briefly described. \label{tab:charac}}}	\vspace{0.5cm}

			%	\begin{tabularx}{0.9916\textwidth}
\centering
	\begin{tabular}{|l|g|g|g|g|g|}
		\hline
		\rowcolor{white}
		& In-degrees & Out-degrees  & Hubs & Constructional principle\\
		\hline
		\textit{Regular network} & constant & constant   & no & fixed number of neighbours \\
		\rowcolor{white}
		SII \textit{random network} ~\cite{Izhikevich.2006} & Poisson & constant   & very unlikely & fixed number of outputs \\
		ER \textit{random network}~\cite{Erdos.1959} & Poisson & Poisson  & unlikely  & fixed connection probability \\
		\rowcolor{white}
		IC \textit{scale-free network}~\cite{Catanzaro.2005}& power-law & power-law   & yes &  connection probability distribution function\\
		BA \textit{scale-free network}~\cite{Barabasi.2003}& power-law & power-law  & yes & growth and preferential attachment \\
		\hline
	\end{tabular}\twocolumn
\end{table}

As reported in~\cite{Swadlow.1994}, synaptic transmission times of 1 to 20\,ms are realistic in cortex. However, the original Izhikevich model simulation of~\cite{Izhikevich.2003} uses uniform delay times of 1\,ms for all connections.
To increase the biological plausibility, here implemented simulations are based on the code by Izhikevich in 2006~\cite{Izhikevich.2006}, which is an advanced version with plasticity obeying the Hebbian theory and axonal conduction delays. The ability of plasticity is called \ac{STDP}, where synchrony dramatically decreases after seconds and network bursts do not appear any more. 
In this approach a static topology is more desirable. Therefore \ac{STDP} is not used, but transmission times randomly distributed with values between 1 and 20\,ms.

%\subsection*{Construction of networks}
\subsection*{Construction and weighing of synapses}\vspace{0.1cm}
For comparison of possible network topologies, their characteristics are described in Tab.~\ref{tab:charac}. The \textit{regular network} was not used for simulation because of its unrealistic construction in nature. The \ac{SII}~\cite{Izhikevich.2006} for \textit{random networks} was constructed with randomly chosen 100 of outgoing synapses for each neuron, where no inhibitory-to-inhibitory connections are allowed. In this way, for a network of 1000 neurons, the resulting connection probability of outgoing synapses is set to $\frac{100}{1000}=0.1$ without standard deviation. For incoming synapses, the probability is also 0.1 with standard deviation.
Furthermore, a second \textit{random network} will be evaluated in form of an implemented \ac{ER} network topology~\cite{Erdos.1959} with a connection probability of 0.1, following a Poisson distribution for input- and output-degrees. Another difference to the \ac{SII} \textit{random network} is the possibility of connections between inhibitory neurons.
Thus, both \textit{random networks} have a theoretical mean degree of 200 synapses: 100 inputs and 100 outputs.\vspace{0.1cm}

Moreover, two ways of \textit{scale-free} construction are investigated. First, the \ac{IC}~\cite{Catanzaro.2005} for uncorrelated random \textit{scale-free networks} uses a connection probability function in form of formula~(\ref{equ:sf_d}). The parameters applied are a minimum degree of 10 and $\gamma = 2.0$. Second, the \ac{BA}~\cite{Barabasi.2003} network, which is also a form of \textit{scale-free network}, with 24 connections, 12 input- and 12 output-synapses, per each growing step were constructed. Thus, the minimum degree of \ac{BA} networks is 24. Inhibitory to inhibitory connections are permitted for both \textit{scale-free network} types.\vspace{0.1cm}

For all networks, self-connections and parallel synapses are prohibited, while antiparallel synapses are allowed.
Binary masks for respective \acp{SWM} were constructed by modified implementations of the Python complex network package \textit{NetworkX}~\cite{Hagberg.2008}.
\acp{SWM} were filled up with log-normal distributed synaptic weights with a maximum of 10. The mean synaptic weight is chosen for each simulation in such a way that network bursts appear. It was found that a higher density of connections required a lower average synaptic weight for regular network bursts.
\vspace{4cm}

%\newpage
\subsection*{Simulation}
Discussed \textit{in silico} networks were implemented and simulated with  \textit{2017a MATLAB}, MathWorks, with a modified version of~\cite{Izhikevich.2006}. The 64-bit computer uses an Intel i5-2400 CPU and 8\,GB RAM. Since the calculation time for 1000 neurons is approximately same to the simulated time, the computer configuration can be recommend. For each topology a 1000 neuron network was simulated four times, in each case for one hour.

\section{Results and Discussion}
\begin{figure}[!b]
	\centering
	\includegraphics[width=0.4\textwidth]{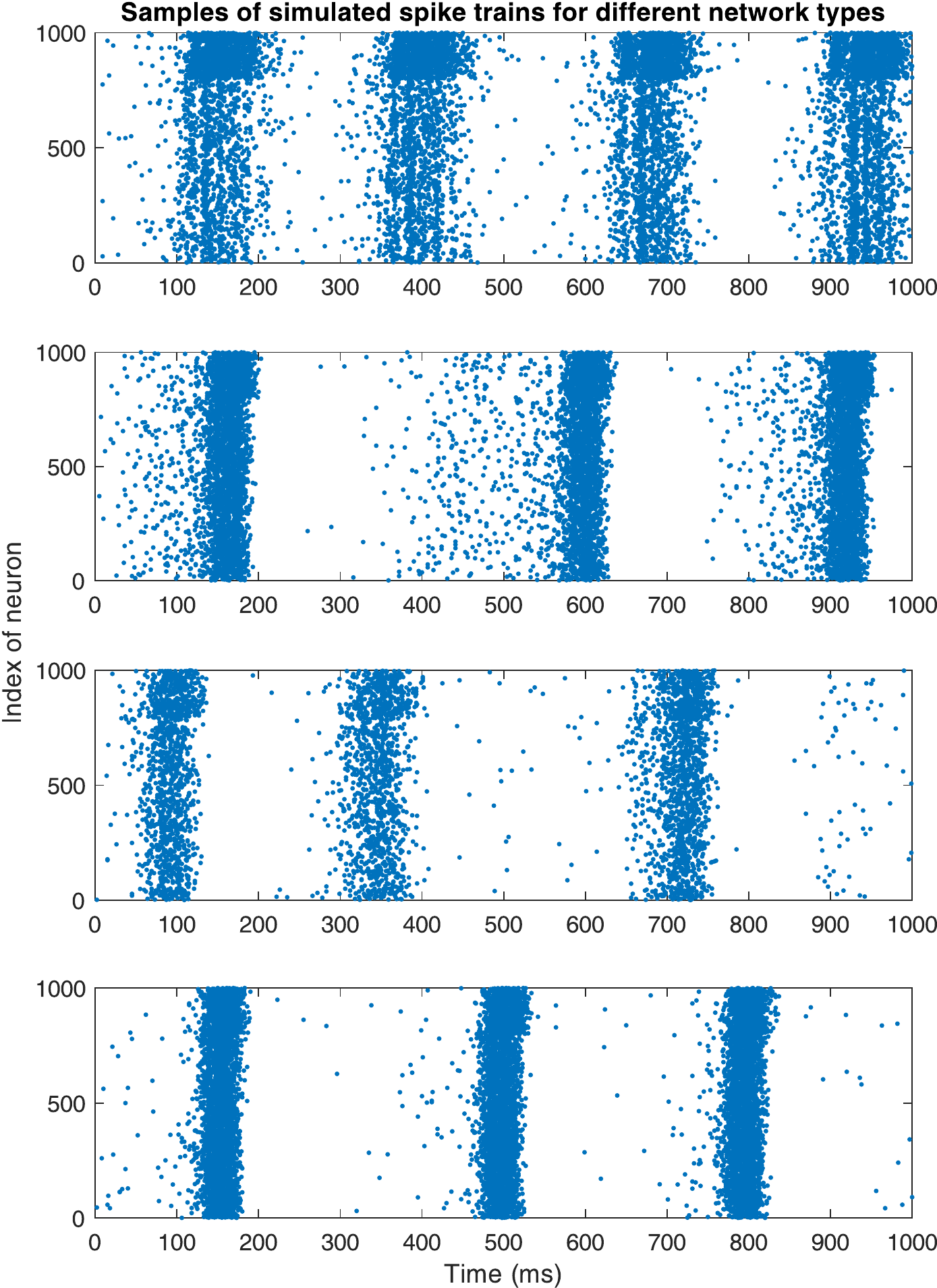}
	\caption[Samples of simulated spike trains]{\textbf{Samples of simulated spike trains:\\}  Each dot represents an emitted action potential of the neuron with index of respective y-coordinate. The networks with overall 1000 neurons each produced three or four network bursts in the shown second of simulation. The samples are sorted for network types from top to bottom: \ac{SII} \textit{random network}, \ac{ER} \textit{random network}, \ac{IC} \textit{scale-free network} and \ac{BA} network. \\}\label{fig:sample}
\end{figure}

Samples of the simulated spike train data are shown in Fig.~\ref{fig:sample}. Network bursts were ensured for all simulations with manual regulation of synaptic weights. In this way, all samples of the explored network types show similar patterns, even if the spiking density varies in network bursts.
The resulting distributions for input-, output- and total-degrees, separated into excitatory and inhibitory neurons, are analysed. Furthermore, respective \acp{MFR} are measured by counting all spikes of neuron $i$, whose history is called spike train $S_{i(t)}$, and dividing the number by the considered time range, see formula (\ref{equ:MFR}).

\begin{equation}\label{equ:MFR}
MFR_{i} =  \frac{\sum_{t=t_{start}}^{t_{end}} S_{i(t)}}{t_{end} - t_{start}}
\end{equation}

\begin{figure}[!t]
	\centering
	\includegraphics[height=0.4\textheight]{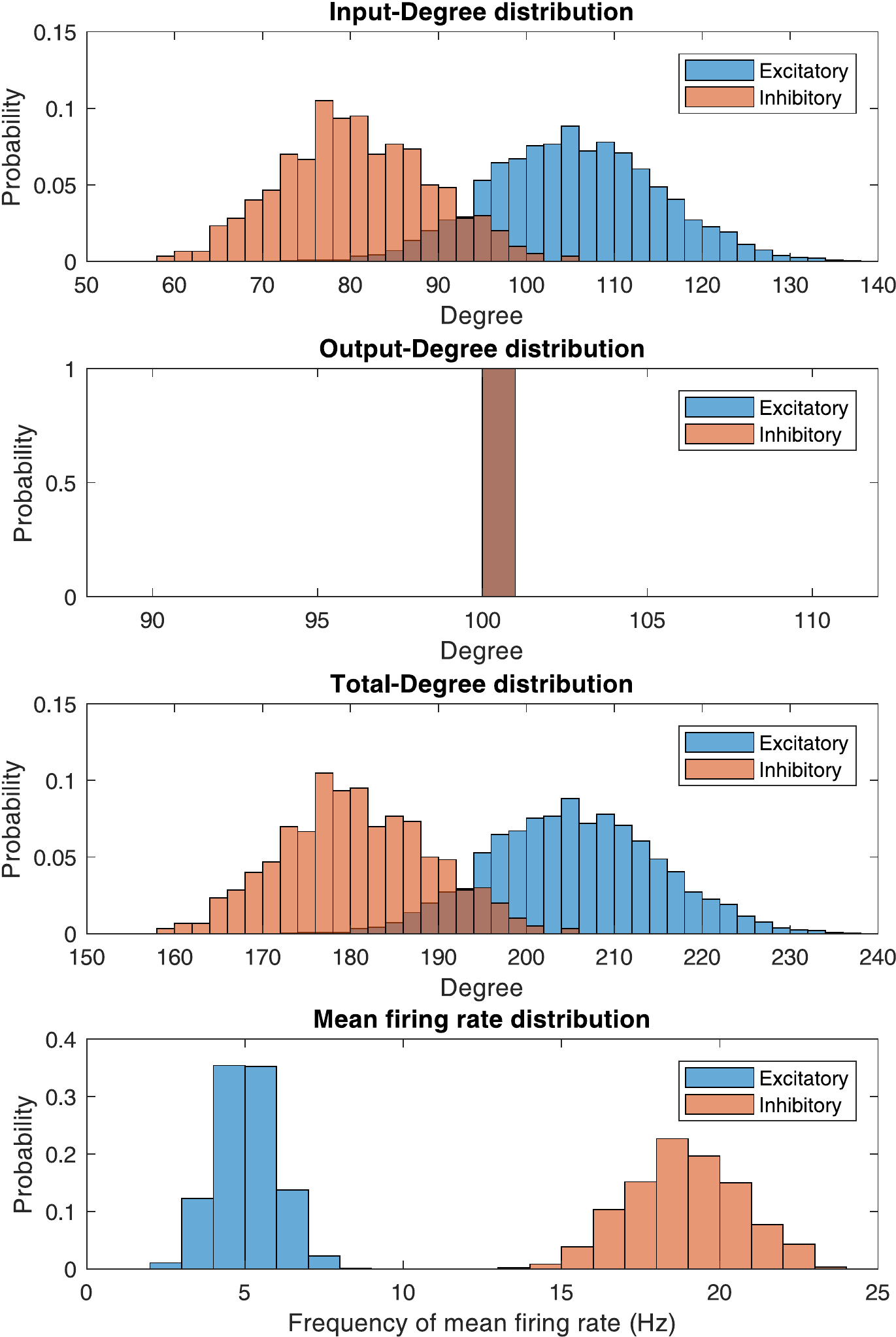}
	\caption[Izhikevich]{\textbf{Analysis of \ac{SII} \textit{random networks}:\\} A lower mean degree of inhibitory neurons is the result of prohibited inhibitory-to-inhibitory connections. The inhibitory effects lead to lower \acp{MFR} of excitatory neurons, while inhibitory neurons will not be inhibited.}\label{fig:IZ}
\end{figure}

Due to the constant output-degrees of 100 for \ac{SII} \textit{random networks}, the difference between its normal distributions of total-degree and input-degree is an offset by 100. Since every fifth neuron is an inhibitory neuron and inhibitory-to-inhibitory connections are not allowed for \ac{SII}, the offset between input-degrees of both neuron types is theoretically $\frac{100}{5}=20$, which is same for the total-degrees. The theoretical assumption can be confirmed by the results, see Fig.~\ref{fig:IZ}, as the mean values of the total degree distributions differ by about 22. It is also the reason for a separation of normal distributed \acp{MFR} for both neuron types. The inhibitory effects lead to inhibited \acp{MFR} of excitatory neurons, while \acp{MFR} of inhibitory neurons are not inhibited. Knowing the effects, in the upper plot of Fig.~\ref{fig:sample} the indexes of inhibitory neurons can be identified easily at indices 801 to 1000.

\begin{figure}[!b]
	\centering
	\includegraphics[height=0.4\textheight]{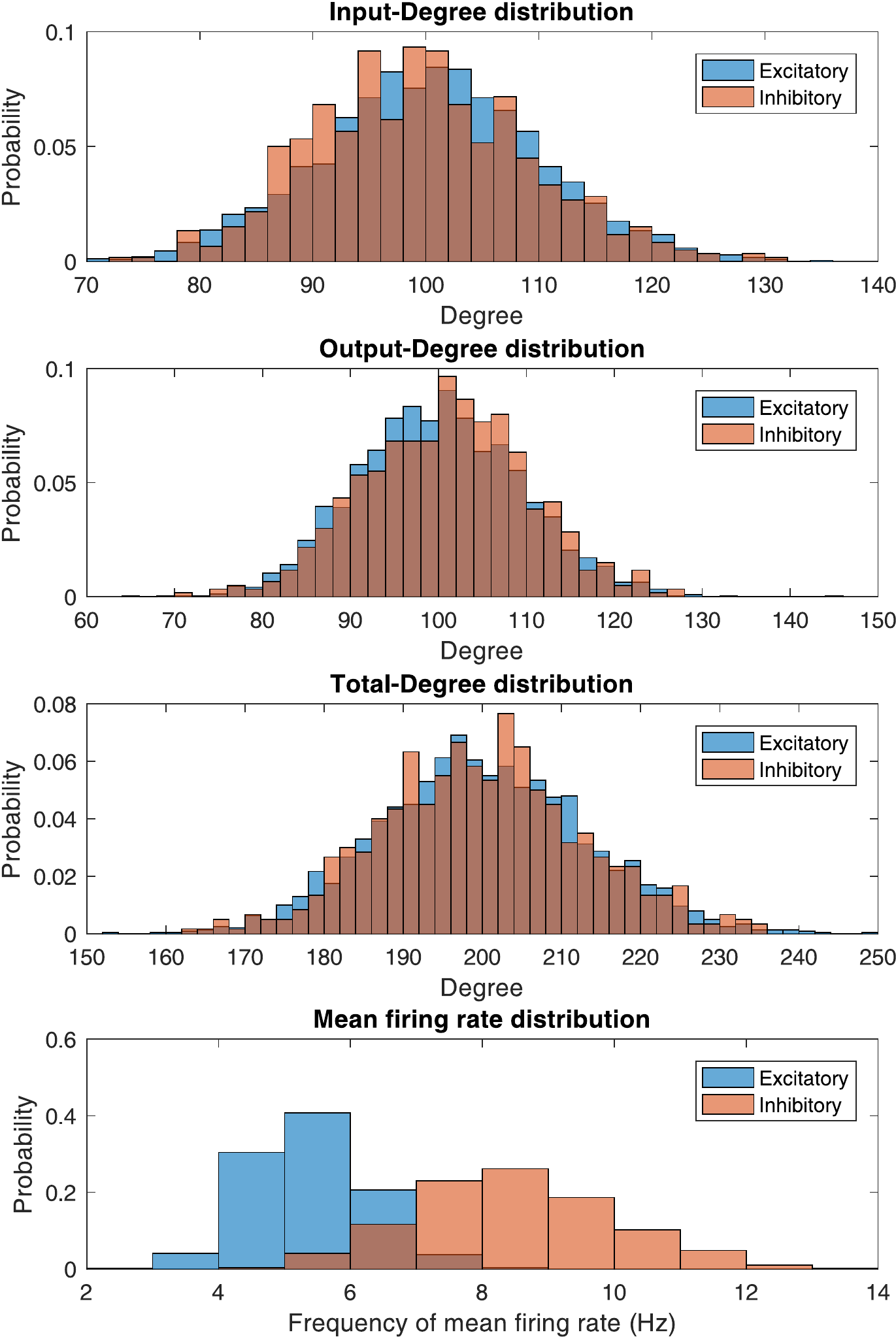}
	\caption[ER]{\textbf{Analysis of generated \ac{ER} \textit{random networks}:\\} In contrast to \ac{SII} \textit{random network}, distribution of \acp{MFR} for both neuron types have a superimposed area.}\label{fig:ER}
\end{figure}

The \ac{ER} model reduces this separation with a non-constant output-degree and the permission for inhibitory-to-inhibitory connections, see Fig.~\ref{fig:ER}. Its normal distributions of input-degree and output-degree are similar to each other with a mean value of 100. In this way the normal distribution of total-degrees has a $\sqrt{2}$ higher standard deviation. Despite superimposed area for \acp{MFR} of both neuron types, the distribution of \acp{MFR} of inhibitory neurons still has a higher mean value with higher standard deviation.
 
\begin{figure}[!h]
	\centering
	\includegraphics[height=0.4\textheight]{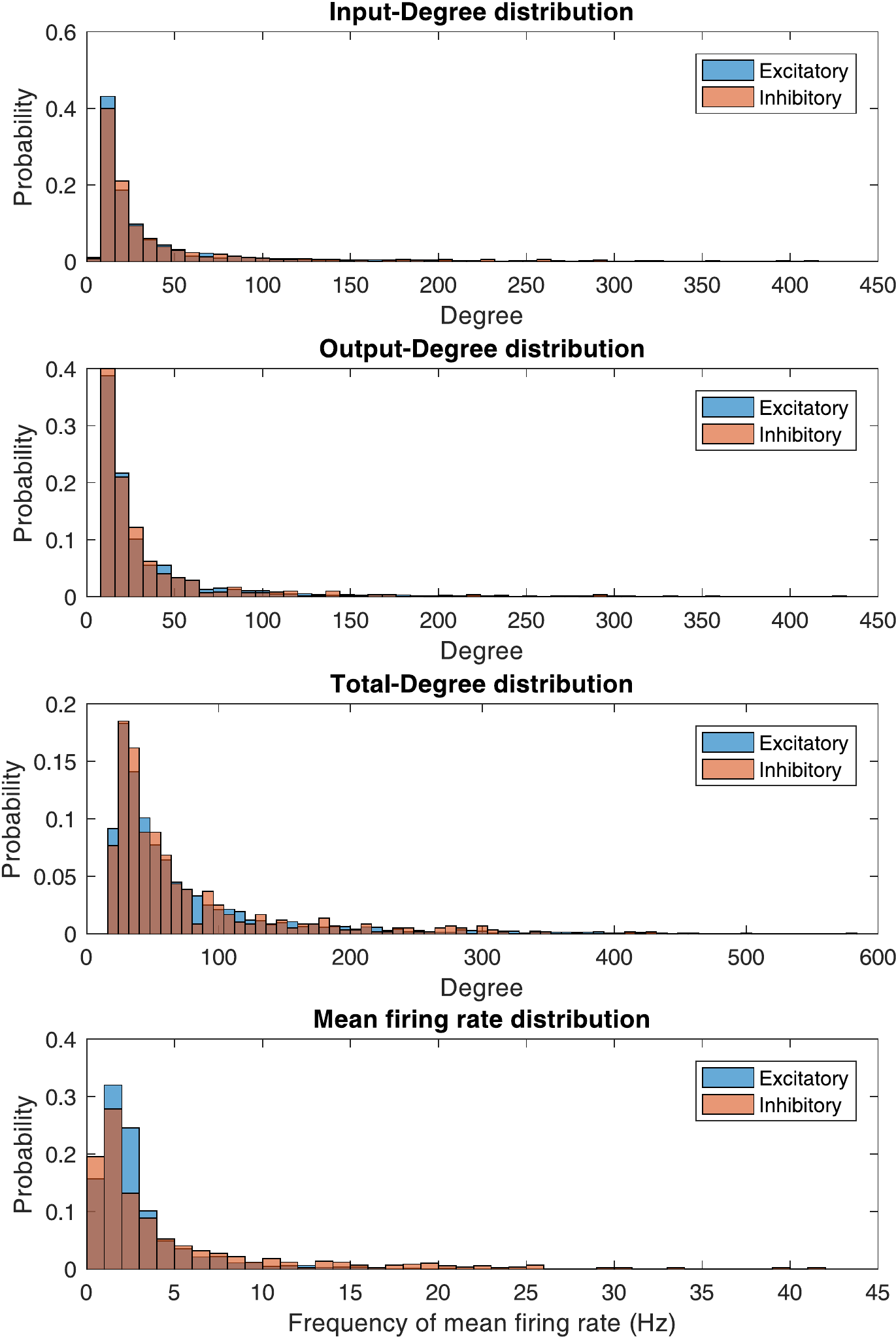}
	\caption[CA]{\textbf{Analysis of generated \ac{IC} \textit{scale-free networks}:\\} Resulting \ac{MFR} distributions of excitatory and inhibitory neurons are similar and can be fitted by a power law function.}\label{fig:CA}
\end{figure}

\begin{figure}[!b]
	\centering
	\includegraphics[height=0.4\textheight]{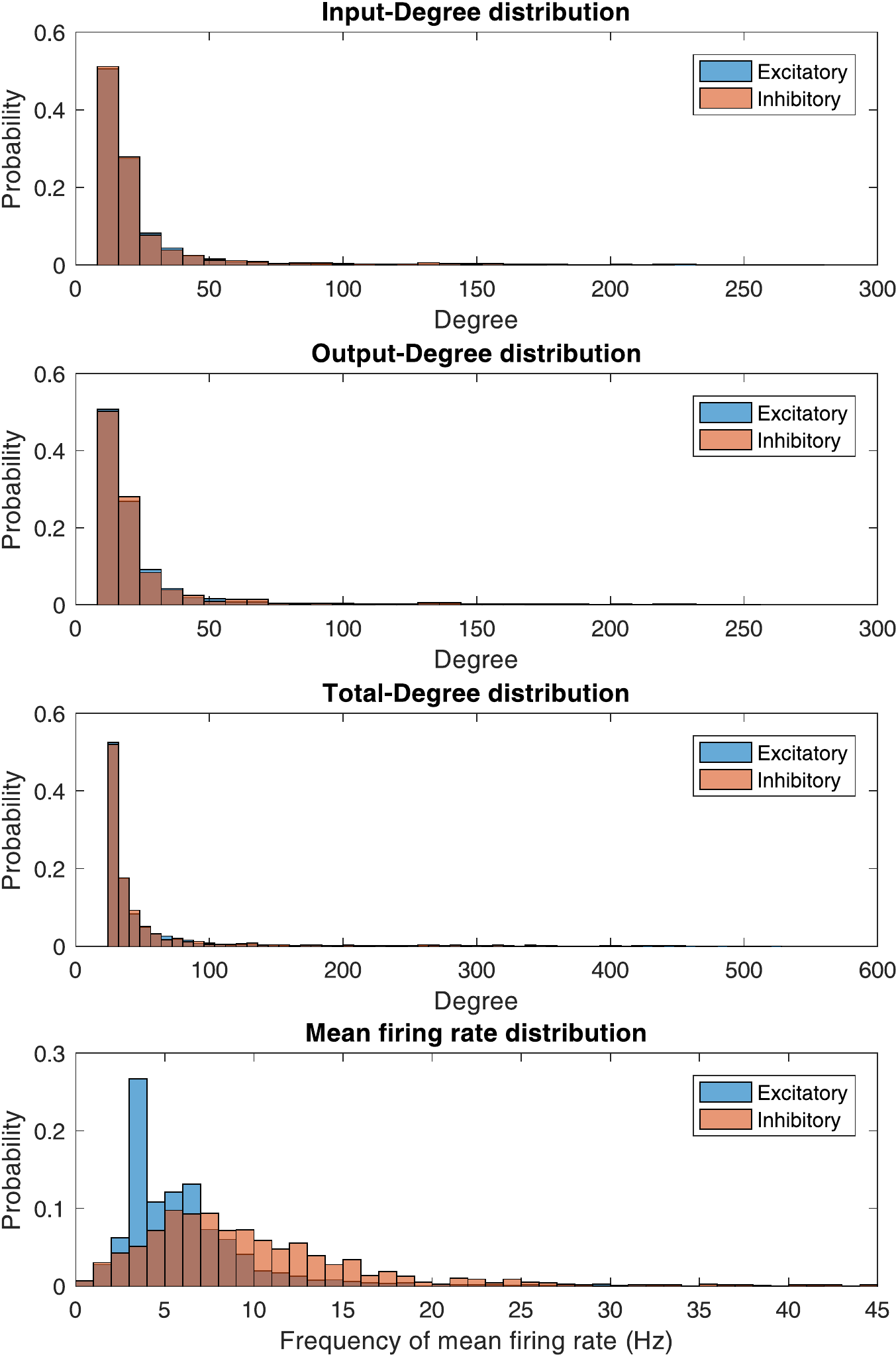}
	\caption[Izhikevich]{\textbf{Analysis of generated \ac{BA} \textit{scale-free networks}:\\} 
	Log normal distributed \acp{MFR} with different standard deviations for both types have a superimposed area.

}\label{fig:BA}
\end{figure}

Both \textit{scale-free} implementations lead to intersections of inhibitory and excitatory \acp{MFR}. For the \ac{IC} networks, log normal distributions of input- and output-degree are similar with a maximal probability at low degrees and outliers can be found at up to 400 connections, see Fig.~\ref{fig:CA}. Thus, total-degrees are also log-normal distributed. Resulting \ac{MFR} distributions of excitatory and inhibitory neurons are similar and can be fitted by a power law function. For the \ac{BA} networks, log normal distributions of input-, output- and total-degree are almost identical for inhibitory and excitatory neurons. The minimum input- and output-degrees are added up to minimum total-degrees. The \ac{BA} network leads to a larger deviation in the distribution of inhibitory \acp{MFR} than that of excitatory \acp{MFR}, see Fig.~\ref{fig:BA}. Respective outliers can be identified at up to 300 inputs or outputs. The total-degree for hub neurons are up to 600. Even with a big superimposed area, the log normal distribution of \acp{MFR} for inhibitory neurons has a higher standard deviation than the one for excitatory neurons.

Summarizing, the distribution of \acp{MFR} can be immensely influenced by modelling different network topologies.
Since a log normal distribution for intracortical spontaneous \acp{MFR} was already demonstrated~\cite{Song.2005}, many neurons fire far above network average in biological \textit{in vitro} cultures. One can conclude that more realistic \acp{MFR} can be measured by using \textit{scale-free} topologies for large scale neural network simulations.

\newpage

\section{Conclusion and Outlook}
For a meaningful evaluation of neurocomputational algorithms analysing large neuro datasets, \textit{in silico} neural networks shall be employed with good biological relevance.
However, the complexity and topology of commonly used \textit{in silico} neural networks and thus the evaluation results can vary.
For example, the accuracy of Transfer Entropy, a widely used algorithm for connectivity estimation, varies greatly depending on the model used for the evaluation~\cite{Garofalo.2009, Pastore.2017}.

The result of this work is indeed that the implemented neural network topology immensely influences the distribution of the \acp{MFR} and thus the meaningfulness of evaluation results. 
Whereas \ac{SII} topologies are used in many evaluations, e.g. without delay times~\cite{Garofalo.2009}, with delay times, and with different use of \ac{STDP}~\cite{Pastore.2017,Ito.2011}, more realistic \textit{scale-free} topologies are rarely used for evaluations~\cite{Kadirvelu.2017}.
One can concluded that searching for the best algorithms can be confusing, since apparently better results are not synonymous with physiologic relevance for real biological neural networks.

For future evaluations, a standardised method improves an effective research of neurocomputational algorithms. Widely used and uniform benchmarking also makes it easier to compare newly developed methods with previous methods.
Moreover, the further development and improvement of intergroup research is possible in a simpler way. To ensure topology independent algorithms, a multiple model evaluation is useful. Usage of at least one \textit{scale-free network} implementation is necessary for good, sufficient biological plausibility. Furthermore, to strengthen the significance repeated simulations are required. Since electrophysiological recording methods do not measure all signals of a neuronal culture, it is recommended to use spike train data of a small subset, e.g. 100 spike trains of a simulated network with 1000 neurons. The use of a common benchmarking with large-scale neural network simulation, taking all these aspects into account, will help to improve research in the field of computational neuroscience.

In order to establish standardised evaluation methods in computational neuroscience, we will continue our research and evaluate several connectivity estimation algorithms with a simulation framework for large scale neural networks with different topologies and make this framework available to all interested parties.

\section*{Acknowledgements}
Research described in the paper was supervised by Prof.\,Dr.-Ing.\,\,C.\,Thielemann and M.\,Eng.\,\,M.\,Ciba, Biomems lab, \ac{UAS} Aschaffenburg. This work was funded by the BMBF Germany within the project Neurointerface (F{\"o}rderkennzeichen 03FH061PX3).

%----------------------------------------------------------
%               THIS IS THE PLACE FOR REFERENCES

\vspace{-0.8cm}
\begin{authorcv}~\end{authorcv}
 \hspace{-0.65cm}
	\begin{tabular}{cl}  
		\begin{tabular}{c}
			\includegraphics[width=3cm]{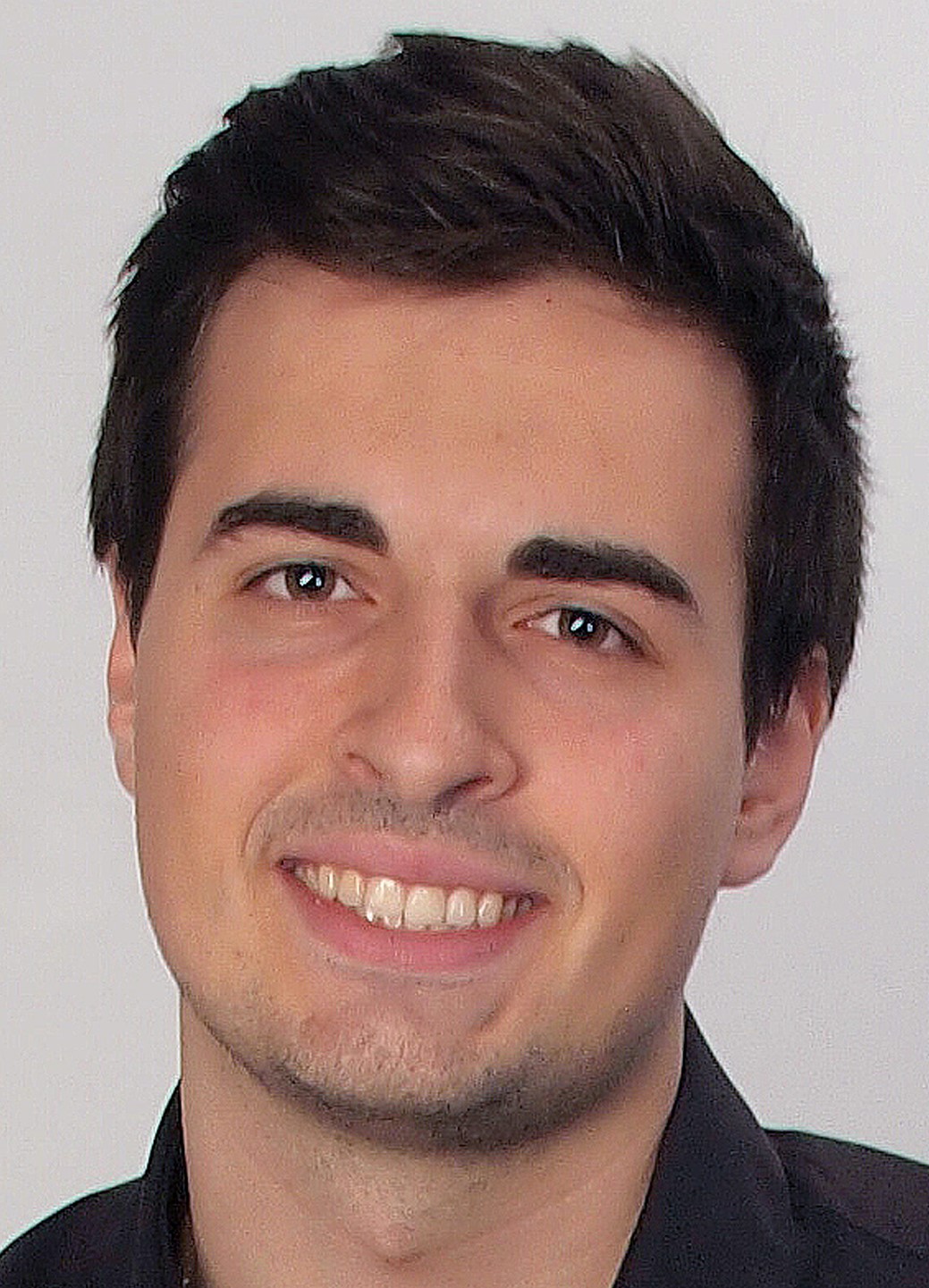}
		\end{tabular}
	 \begin{tabular}{l}
			\parbox{0.6\linewidth}{
				\textbf{Stefano DE BLASI} was born in Erlenbach am Main, Germany. After working as an electronics technician, he graduated with a bachelor's degree in Electrical Engineering and Information Technology at \ac{UAS} Aschaffenburg. Currently he studies for a master's degree. His research focuses on computational neuroscience.
			}
		\end{tabular}  \\
	\end{tabular}

\end{document}